\documentclass[manuscript]{emulateapj}

\usepackage{graphicx}
\usepackage[space]{grffile}
\usepackage{latexsym}
\usepackage{amsfonts,amsmath,amssymb}
\usepackage{url}
\usepackage[utf8]{inputenc}
\usepackage{fancyref}
\usepackage{hyperref}
\hypersetup{colorlinks=false,pdfborder={0 0 0},}
\usepackage{gensymb}

\usepackage{natbib}

\def\rsun{\ifmmode {\rm R}_{\mathord\odot}\else $R_{\mathord\odot}$\fi}
\def\msun{\ifmmode {\rm M}_{\mathord\odot}\else $M_{\mathord\odot}$\fi}
\def\lsun{\ifmmode {\rm L}_{\mathord\odot}\else $L_{\mathord\odot}$\fi}
\def\kms{\ifmmode {\rm km ~ s}^{-1}\else km s$^{-1}$\fi}
\shorttitle{Predictions for Observing Protostellar Outflows with ALMA}
\shortauthors{Bradshaw et al.}

\begin{document}


\title{Predictions for Observing Protostellar Outflows with ALMA}


\author{Christopher Bradshaw}\affil{Department of Astronomy,  Yale University, New Haven, CT 06511}

\author{Stella S. R. Offner}\affil{Department of Astronomy, Yale University, New Haven, CT 06511\footnote{Hubble Fellow}; Department of Astronomy, University of Massachusetts, Amherst, MA 01003.} 

\author{H\'ector G. Arce}
\affil{Department of Astronomy, Yale University, New Haven, CT 06511}

\begin{abstract}
Protostellar outflows provide a means to probe the accretion process of forming stars and their ability to inject energy into their surroundings. However, conclusions based on outflow observations depend upon the degree of accuracy with which their properties can be estimated. We examine the quality of ALMA observations of protostellar outflows by producing synthetic $^{12}$CO(1-0) and $^{13}$CO(1-0) observations of numerical simulations.
 We use various ALMA configurations, observational parameters, and outflow inclinations to assess how accurately different assumptions and setups can recover underlying properties. We find that more compact
arrays and longer observing times can improve the mass and momentum recovery by a factor of two.  During the first $\sim$0.3 Myr of evolution, $^{12}$CO(1-0) is optically thick, even for velocities $|v|\ge 1 \kms$,  and outflow mass is severely underestimated without an optical depth correction. Likewise, $^{13}$CO(1-0) is optically thick during the first $\simeq 0.1$ Myr.  However, underestimation due to shorter observing time, missing flux, and optical depth are partially offset by the assumption of LTE and higher excitation temperatures. Overall, we expect that full ALMA $^{13}$CO(1-0) observations of protostellar sources within 500 pc  with observing times $\gtrsim 1$ hrs and assumed excitation temperatures of $T<20$K will reliably measure mass and line-of-sight momentum to within 20\%.
\end{abstract}


%\keywords{globular clusters: general --- globular clusters: individual(NGC 6397, NGC 6624, NGC 7078, Terzan 8}

\bibliographystyle{apj}

\section{Introduction}

Young protostars are observed to launch energetic collimated bipolar mass outflows \citep{lada85, bachiller96}. These protostellar outflows play a fundamental role in the star formation process on a variety of scales. On sub-pc scales they entrain and unbind core gas, thus setting the efficiency at which dense gas turns into stars \citep{matzner99,alves07,machida13,Offner14b}.  Interaction between outflows and infalling material may  regulate protostellar accretion and, ultimately, terminate it \citep{arce07,frank14}. On sub-pc up to cloud scales, outflows inject substantial energy into their surroundings,  potentially providing a means of sustaining cloud turbulence over multiple dynamical times \citep{nakamura07,carroll09,wang10,arce10,nakamura11,hansen12,nakamura14}.

The origin of outflows is attributed to the presence of magnetic fields, and a variety of different models have been proposed to explain the launching mechanism. Of these, the ``disk-wind" model \citep{blandford82,pelletier92}, in which the gas is centrifugally accelerated from the accretion disk surface, and the ``X-wind" model \citep{shu88}, in which gas is accelerated along tightly wound field lines, are most commonly invoked to explain observed outflow signatures. However, investigating the launching mechanism is challenging because launching occurs on scales of a few stellar radii and during times when the protostar is heavily extincted by its natal gas.  Consequently, separating outflow gas from accreting core gas, discriminating between models, and determining fundamental outflow properties are nontrivial.

%Three main approaches have been applied to studying outflows. 
Single-dish molecular line observations have been successful in mapping the extent of outflows and their kinematics on core to cloud scales \citep[e.g.,][]{lada96,tafalla97,bourke97,bally99,yu99,stojimirovic06,curtis10,arce10,dunham14}. However, outflow gas with velocities comparable to the cloud turbulent velocity can only be extracted with additional assumptions and modeling \citep[e.g.,][]{bontemps96,arce01b,maury09,dunham14}, which are difficult to apply to confused, clustered star forming environments \citep[e.g.,][]{curtis10,graves10,plunkett13}. Meanwhile, small scale structure and the inner launching region, which can provide clues about entrainment and variability, are unresolved.
%Interferometry provide a means of mapping outflows down to scales of 10$^2$ to 10$^3$ AU scales  \citep[e.g.,][]{gueth99,lee02,arce06}, and Early Science observations with the Atacama Large Millimeter/submilllimeter Antenna (ALMA) have extended these limits down to resolutions of nearly 100 AU  \citep{codella14,lee14}. 
Most interferometer observations have provided outflow maps with resolution of 10$^2$ to 10$^3$ AU scales  \citep[e.g.,][]{gueth99,lee02,arce06,hirano10}, but a few recent interferometer maps, including the latest maps obtained with the 
Atacama Large Millimeter/submilllimeter Antenna (ALMA) can probe down to scales of tens to about 100 AU \citep{codella07,lee09,cabrit12,codella14,lee14,oya14}
Interferometry is sometimes able to produce large high-resolution maps \citep{plunkett13,fernandezlopez14,storm14}, but it can also resolve out larger scale structure if short-spacing or single-dish observations are not combined with the interferometer data. Consequently, it is sometimes difficult to assemble a complete and multi-scale picture of outflow properties with these observations. Uncertain optical depth, variation in excitation and abundance, and projection effects further complicate the extraction of physical information.

In the absence of either multiple emission lines or tracers, gas above a few km s$^{-1}$ is often assumed to be optically thin  and in LTE \citep[e.g.][]{cabrit90,lada96,bourke97,dunham14}. Analytic models provide a means to evaluate these assumptions, however most models assume outflow symmetry, smooth velocity and density distributions, and a simple temperature profile  \citep[e.g.,][]{canto91,masson93,raga93,canto00,lee01,carolan08}.
In contrast, three-dimensional, hydrodynamic simulations supply detailed predictions for physical quantities related to launching, entrainment and energy injection \citep{seifried12,machida13,Offner14b}, which allow observational methods to be tested rigorously.
%launching, entrainment and energy injection \citep{seifried12,machida13,Offner14b}. 
Thus, the most promising avenue for constructing a complete picture of outflows lies at the intersection of numerical modeling and observations. By performing synthetic observations to model molecular and atomic lines, continuum, and observational effects, simulations can be mapped into the observational domain where they can be compared directly to observations \citep[e.g.,][]{Offner11,Offner12b,Mairs14}. Such direct comparisons are important for assessing the ``reality" of the simulations,  to interpret observational data and to assess observational uncertainties \citep{goodman11}. In addition to observational instrument limitations, chemistry and radiative transfer introduce additional uncertainties that are difficult to quantify without realistic models \citep{beaumont13}. Synthetic observations have previously been performed in the context of understanding outflow opening angles \citep{Offner11}, observed morphology \citep{peters14}, and impact on spectral energy distributions \citep{Offner12}. 

%However,  numerical simulations may inhabit a very different parameter space from the observational data. The most promising means of investigation involves producing synthetic observations of the numerical data.
%Henceforth we use outflow and jet interchangeably.
The immanent completion of ALMA provides further motivation for predictive synthetic observations. Although ALMA will have unprecedented sensitivity and resolution compared to existing instruments, by nature interferometry resolves out large-scale structure and different configurations will be sensitive to different scales. Atmospheric noise and  total observing time may also effect the fidelity of the data. Previous synthetic observations performed by \citet{Offner12b} suggest that the superior resolution of full ALMA and the Atacama Compact Array (ACA) will be able to resolve core structure and fragmentation prior to binary formation. \citet{peters14} predict that ALMA will be able to resolve  complex outflow velocity structure and helical structure in molecular emission. However, neither of these studies was sufficiently comprehensive to assess the fidelity of ALMA observations under different assumptions and conditions.

In this paper we seek to quantify the accuracy of various ALMA configurations in recovering fundamental molecular outflow properties such as mass, line-of-sight momentum, and energy. We use the {\sc casa} software package to synthetically observe protostellar outflows in the radiation-hydrodynamic calculations of \citet{Offner14b}. These simulations supply realistic distributions for gas densities, temperatures, and velocities, which more closely represent the complexity of observed outflows than analytic models. By modeling the emission at different times, inclinations, molecular lines, and observing configurations we evaluate how well physical quantities can be measured in the star formation process. In section \S\ref{methods} we describe our methods for modeling and observing outflows. In section \S\ref{results} we evaluate the effects of different observational parameters on bulk quantities. We discuss results and summarize conclusions in \S\ref{conclusions}.

\section{Methods}\label{methods}

\subsection{Hydrodynamic Simulation}

In order to assess the recovery of information using ALMA, we post-process a self-gravitating radiation-hydrodynamic simulation in which we have complete three-dimensional temperature, density and velocity information. The simulation we focus on here,  th0.1fw0.3, was previously presented in \citep[][hereafter OA14]{Offner14b}. This calculation is not intended to model a specific source but rather to represent a ``generic" molecular outflow originating from a low-mass protostellar source. The outflow is assumed to be driven by an X-wind with a collimation angle that results in a high-momentum region along the poles (which mimics a jet) and a lower momentum wider angle component. Since the simulation follows the evolution over 0.5 Myr it could in principle be used to model outflows from young protostars (e.g., L1448, HH 212, HH 211) as well as slightly more evolved protostars (e.g., HH46/47, L1551-IRS5, L1228). We briefly describe the simulation properties below and refer the reader to OA14 for additional details. 

The calculation was performed with the {\sc orion} adaptive mesh refinement  (AMR) code \cite{truelove98,klein99}. The simulation follows the collapse of an isolated, turbulent low-mass core. It begins with an initially uniform, cold $4 M_\odot$ sphere of radius $R_c=2\times 10^{17}$cm, density $\rho_c=2\times 10^{-19}$ g cm$^{-3}$ and  temperature $T_c=10$ K. This core is embedded in a warm, diffuse gas with $\rho=\rho_c/100$ and $T=100 T_c$ = 1000 K. The dense gas is initialized with a grid of random velocity perturbations such that the initial rms velocity dispersion is 0.5 km s$^{-1}$.  

Additional levels of adaptive mesh refinement (AMR) are inserted as the core collapses under the influence of gravity. The core itself is resolved  with a minimum cell size of $\Delta_{\rm min}\simeq 0.001$ pc, where the maximum level of refinement has $\Delta_{\rm min}\simeq$ 26 AU. Once the central region exceeds the maximum grid resolution ($\rho_{\rm max}\simeq 6.5 \times 10^{-15}$ g cm$^{-3}$, \citep[e.g.,][]{truelove98}),  a ``star" forms. This star, which is represented by a Lagrangian sink particle, accretes, radiates and launches a collimated bipolar outflow \citep{krumholz04,Offner09,cunningham11}.  The rate of mass loss due to the outflow is set to a fixed fraction of the instantaneous accretion rate: $\dot m_w = f_w \dot m_*$, where $f_w = 0.3$ is the outflow launching rate given by the X-wind model \citep{shu88}. The distribution of outflow momentum  is parameterized by a fixed collimation angle, $\theta_0=0.1$, which is empirically determined to be similar to that of observed outflows \citep{matzner99,cunningham11}.  Although $\theta_0$ is constant in time, the outflow injection into the AMR grid occurs on such small scales that the outflow properties such as the opening angle and morphology evolve hydrodynamically. Previous work by \citet{Offner11} and \citet{Offner14b} show that these properties agree well with observed low-mass outflows.  In particular, the outflow morphologies characteristic of CO emission appear similar to those of observed outflows \citep[e.g.,][]{lee02}and the simulations reproduce the outflow opening angle dependence on time seen by \citet{arce06}.

In the simulation the protostar forms around $t\simeq 0.17$ Myr, with the outflow launching beginning shortly thereafter (when the protostellar mass reaches 0.05 $\msun$). For the first few kyr the outflow is confined to the core. It succeeds in breaking out around  $t\sim $0.22 Myr, at which point the gas exhibits well-defined ``v-shaped" cavity walls. After breakout, the outflow entrains and ejects a significant fraction of the core, sweeping up progressively wider solid angles of gas. By $t=0.35$ Myr, much of the core has been dispersed and the outflow structure is no longer clearly visible in the dense gas. Most of the residual gas is either cold and clumpy or  warm and diffuse. The protostar continues accreting from its local accretion disk, but infall has significantly diminished. The outflow continues but at two orders of magnitude below the initial launching rate and little gas is entrained. As a result, the outflowing gas does not exhibit a clear signature in the cold (molecular) gas.
 
By the end of the calculation ($t= 0.5$ Myr), the simulation contains a single star with a mass of $\sim$1.45 $\msun$. 
%Over the evolution, the outflow unbinds and ejects a significant fraction of the gas from the domain, producing a star formation efficiency of 47\%, which is comparable to observational and theoretical estimates for dense cores (e.g., \citealt{matzner00,alves07}).  
Most of the core mass has been ejected from the domain, and the remaining gas has a rms mass-weighted velocity dispersion of $\sim$1 km s$^{-1}$. 

Since OA14 found that the final stellar mass and star formation efficiency did not depend strongly on $\theta_0$ and $f_w$, we analyze only a single calculation. However, we note that different initial core masses, rotations, and magnetic field strengths might produce qualitatively different results \citep{machida13}.

In the remainder of the paper, we focus on the colder molecular component of the simulated outflow, as would be traced by low-level rotational transitions of CO, rather than on the warmer jet component, which is predominantly atomic or ionized gas here. The highest velocity gas (the jet) in the simulation reaches about 75 \kms, which is similar to the deprojected flow speeds estimated for low-mass sources like HH212 \citep{lee08,cabrit12}. In contrast, the cold, molecular component has much lower average velocities of a few \kms.

\subsection{Molecular Line Modeling}

We use the non-local thermodynamical equilibrium radiative transfer code {\sc radmc-3d}\footnote{http://www.ita.uni-heidelberg.de/~dullemond/software/radmc-3d/} to compute the line emission 
in $^{12}$CO(1-0) and $^{13}$CO(1-0). We adopt the Large Velocity Gradient (LVG) approximation \citep{shetty11}, which solves for the rotational level populations by solving the equations for local radiative statistical equilibrium. {\sc radmc-3d} requires 3D input gas densities, velocities and temperatures, which are produced as outputs by the hydrodynamic simulation.  We perform the radiative transfer on a uniform $256^3$ grid, where we interpolate all the AMR data to the second refinement level ($\Delta x =0.001$pc). We include turbulent line broadening on scales at and below the grid resolution by adding a constant micro-turbulence of 0.05 km s$^{-1}$. 
%SSRO took out
%For $^{12}$CO we smooth the velocity field by using a doppler parameter of 0.025, which ensures that the velocity field is linearly interpolated between velocity jumps greater than 0.025$c_s$, where $c_s$ is the local sound speed. $^{13}$CO has a doppler parameter of 0.25. 

%{\bf Why the factor of 10 difference in this parameter between the two isotopologues?}
%This parameter mainly affects the emission in cells abutting the warm atomic gas; it has negligible impact on the emission of the cooler, denser gas.
%doppler parameter--interpolates gas velocity when computing the interaction between the photon and gas. (detailed in documentation).
% Chris: Not doing anything about this as reviewer didn't comment on it, and most people don't mention it.

To obtain the CO abundances from the total gas density, we assume that molecular Hydrogen dominates in all gas cooler than 1,000 K, where $n_{{\rm H}_2}=\rho/(2.8 m_p)$.  We adopt constant CO abundances of [$^{12}$CO/H$_2$]=$8.6 \times 10^{-5}$ \citep{frerking82} and [$^{12}$CO/$^{13}$CO]=62 {\citep{langer93} for gas cooler than 900 K; otherwise the CO abundance is set to zero. Thus,  we adopt a simple prescription for the CO abundance and do not follow chemical networks (e.g. \citealt{viti04,Offner14}). Consequently, the CO line emission only originates from relatively cold gas in the dense core and gas entrained by the outflow;  the warm, low-density ambient material and the hot, outflow gas, which is ionized by construction, do not emit. We adopt the molecular collisional coefficients from \citet{schoier05}. The output spectral cube has a velocity resolution of 0.08 \kms and velocity range $|v| \le 10$ \kms.  Little CO emission originates from gas moving faster than 10 \kms, because the simulated gas tends to be atomic or ionized. As cold material is swept up by the outflow, the net gas velocity diminishes. Observed outflows often have more emission at higher velocities than we find here \citep[e.g.][]{bachiller91,bachiller95,cernicharo96,gueth99,vandermarel13,dunham14}. The difference could arise because the adopted launching model underestimates the amount of high-velocity gas or because the entrainment of cold gas is even more efficient than predicted by the simulations.

\subsection{Interferometry Modeling}

We convert the {\sc radmc-3d} outputs into skymaps in units of Janzky px$^{-1}$ that we post-process using the  Common Astronomy Software Applications (CASA)\footnote{\href{http://casa.nrao.edu}{http://casa.nrao.edu}}.  For our fiducial parameters, we adopt the distance, $d=450$ pc, and sky position, $(\alpha, \delta)=08^h25^m41 5, -51{\arcdeg}00'47''$ (J2000),  of HH46/47, an outflow recently observed with ALMA Cycle 0 \citep{arce13}.  The simulation is not intended to be an exact model of HH46/47 but instead represent a typical young, isolated low-mass outflow of which HH46/47 is one example.

We use the CASA task ``simobserve" to produce a model observation for a given antenna configuration and set of observing conditions. We add thermal noise assuming 0.5 mm of precipitable water vapor and a ground temperature of 269 K, which reflect good observing conditions.   The synthetic maps are comprised of 27 pointings, which cover an area of approximately 128''$\times$128''. We deconvolve and clean the synthetic observation using the task ``simanalyze".  We vary the ALMA configuration, pointing time, cleaning parameters, and integration times as indicated in Table \ref{tab_runs}. We find that the choice of cleaning parameters had little impact on the recovered mass (see the Appendix for further detail), and so we focus our analysis on the other parameters.

%\begin{center}
\begin{deluxetable*}{c  c  c  c  c  c  c  c } 
\tablecolumns{8}
 \tablecaption{CASA Run Parameters  \label{tab_runs}}
%\begin{tabular}{c | c | c | c | c | c | c | c } 
\tablehead{ \colhead{Run\tablenotemark{a}} &  \colhead{CO Isotope}  &   \colhead{$\theta_{\rm los}$} &  \colhead{Configuration}  &  \colhead{$t_{\rm tot}$} &  \colhead{$t_{\rm pointing}$} &  \colhead{$N_{\rm clean}$} &  \colhead{$F_{\rm clean}$ (mJy)}} \\
\startdata
R1  &  12  &  30  &  Compact  &  7200 & 10  &  10000 & 0.1 \\
R2  &  12  &  30  &  Cycle 1.3  & 7200 & 10  &  10000 & 0.1 \\
R3  &  12  &  30  &  Full 3  &  3600 & 10  &  10000 & 0.1 \\
R4  &  12  &  30  &  Full 3  &  7200 & 10  &  10000 & 0.1 \\
R5  &  12  &  45  &  Compact  &  7200 & 10  &  10000 & 0.1 \\
R6  &  12  &  45  &  Cycle 1.1  & 7200 & 10  &  10000 & 0.1 \\
R7  &  12  &  45  &  Full 3  &  3600 & 10  &  10000 & 0.1 \\
R8  &  12  &  45  &  Full 3  &  7200 & 10  &  10000 & 0.1 \\
R9  &  13  &  45  &  Compact  &  7200 & 10  &  10000 & 0.1 \\
R10  & 13  &  45  &  Cycle 1.1  & 7200 & 10  &  10000 & 0.1 \\
R11  &  13  &  45  &  Full 3  &  3600 & 10  &  10000 & 0.1 \\
R12  &  13  &  45  &  Full 3  &  7200 & 10  &  10000 & 0.1 \\
R13  &  13  &  30  &  Cycle 1.1 & 7200 & 30  &  10000 & 0.01 \\
R14  &  13  &  30  &  Cycle 1.1 & 7200 & 30  &  10000 & 0.1 \\
R15  &  13  &  30  &  Cycle 1.1 & 7200 & 30  &  10000 & 1.0 \\
R16  &  13  &  30  &  Cycle 1.3 & 7200 & 10  &  10000 & 0.1 \\
R17  &  13  &  30  &  Cycle 1.3 & 7200 & 30  &  10000 & 0.1 \\
R18  &  13  &  30  &  Compact  &  7200 & 10  &  10000 & 0.1 \\
R19  &  13  &  30  &  Compact  &  7200 & 30  &  10000 & 0.1 \\
R20  &  13  &  30  &  Compact  &  7200 & 100  &  10000 & 0.1 \\
R21  &  13  &  30  &  Full 3  &  3600 & 10  &  1000 & 0.1 \\
R22  &  13  &  30  &  Full 3  &  3600 & 10  &  10000 & 0.1 \\
R23  &  13  &  30  &  Full 3  &  3600 & 10  &  20000 & 0.1 \\
R24  &  13  &  30  &  Full 3  &  300  &  10  &  10000 & 0.1 \\
R25  &  13  &  30  &  Full 3  &  600  &  10  &  10000 & 0.1 \\
R26  &  13  &  30  &  Full 3  &  7200 & 10  &  10000 & 0.1 
\enddata
%\medskip
\tablenotetext{a}{Run name, CO isotopologue, viewing angle with respect to the outflow axis, ALMA antenna configuration, total integration time, pointing time, number of cleaning iterations, and the cleaning threshold.}
\end{deluxetable*}
%\end{center}

\subsection{Derivation of Mass, Momentum and Energy}

It is possible to estimate the mass from the emission in a single CO line if several assumptions are applied \citet{bourke97}:  (1) the gas is in local thermodynamic equilibrium (LTE), (2) the molecular line is optically thin, and (3) the level populations can be modeled with a single excitation temperature.  One or more of these assumptions is routinely made when estimating outflow properties in observations \citep[][and references therein]{dunham14}. When multiple lines or tracers are available a more rigorous analysis can be performed, however,  constraining the accuracy of these approximations without full information is challenging. By analyzing the simulations in which we have complete knowledge of the underlying mass and temperature distribution, we can individually examine these approximations and  quantitatively estimate the errors introduced.

The simplest mass derivation adopts all three assumptions. In this case, the outflow mass derived from a $^{12}$CO voxel with channel velocity $v$ and position $(\alpha, \delta)$ is given by:
\begin{equation} \label{eqn:massfromtemp}
M_{12}(v, \alpha, \delta) = [{\rm H}_2/^{12}{\rm CO}] \mu_m A(\alpha, \delta) F (T_{ex}) T_B(v, \alpha, \delta)\Delta v
\end{equation}
where 
\begin{equation} %\label{eqn:F(T_ex)}
F(T_{\rm ex}) = 2.31 \times 10^{14} \frac{T_{\rm ex} + 0.92}{1 - e^{-5.53/T_{\rm ex}}} \frac{1}{J(T_{\rm ex}) - J(T_{\rm bg})}
\end{equation}
and
\begin{equation} %\label{eqn:J(T)}
J(T) = \frac{h\nu}{k(e ^ {\frac{h\nu}{kT}} - 1)}.
\end{equation}
Here, $[{\rm H}_2/^{12}{\rm CO}]$ is the abundance ratio of H$_2$ to  $^{12}$CO, $\mu_m=2m_p$ is the mean molecular mass of H$_2$ \footnote{ Note that typically, but not always,  $\mu_m$ takes into consideration the typical Helium and metal abundances in the cloud to obtain the total gas mass, in which case it has a value of 2.8$m_p$ for gas composed of 71\% H, 27\% He and 2\% metals, see, e.g., \citealt{dunham14}.}
$A$ is the area of emission ($\Delta x^2$), $T_{\rm ex}$ is the excitation temperature, $T_{\rm bg}$ is the background temperature, $T_B$ the brightness temperature, $v$ is the velocity, $k$ is the Boltzmann constant, $\nu$ is the frequency of the line transition ($\nu_{^{12}{\rm CO(1-0)}}$=115.3 GHz, $\nu_{^{13}{\rm CO(1-0)}}$=110.2 GHz), and $h$ is Planck's constant. 
Likewise, the mass estimated from the $^{13}$CO emission is $M_{\rm 13}(v, \alpha, \delta)=[^{12}{\rm CO}/^{13}{\rm CO}] M_{\rm 13}(v, \alpha, \delta)$, where we adopt an isotopic ratio of $[^{12}{\rm CO}/^{13}{\rm CO}]=62$. In observations, $T_{\rm bg}=2.73$ K, however, we set $T_{\rm bg}=0$ K, since we do not include the cosmic microwave background in the modeling.  In the fiducial case, we adopt an excitation temperature of $T_{\rm ex} = 11$ K, which reflects that the molecular outflow is dominated by entrained material. This is the value adopted by \citet{bourke97}, and we find that 11 K is in fact similar to the typical molecular gas temperature in the simulation (see Table \ref{tab_texcit}). We assess the accuracy of this value in \S\ref{results_texcit}.

We derive the brightness temperature, $T_{B}$, in each voxel from the emission computed by {\sc radmc-3d} assuming the gas is a  blackbody in the Rayleigh-Jeans limit:
\begin{equation} %\label{eqn:Rayleigh jean}
T_B = \frac{ B_{\nu} c^2}{2 \nu^2 k},
\end{equation}
where $B_{\nu}$ is the intensity.
The total mass is then
\begin{equation} %\label{eqn:Mass from temp}
M_{\rm tot }= \int_{v_{\rm min}}^{v_{\rm max}}\int_{\alpha,\delta} M(v, \alpha, \delta) dv d\alpha d\delta,
\end{equation}
where $v_{\rm min}$ and $v_{\rm max}$ are either the expected velocity range of the outflow or the limits of the spectral cube ($\pm10\,\kms$).

To obtain the momentum and energy, we use the estimated mass for each voxel as well as the line-of-sight velocity:
\begin{eqnarray} %\label{eqn:Mass from temp}
P_{\rm tot }&=& \int_{v_{\rm min}}^{v_{\rm max}} \int_{\alpha,\delta}  M(v, \alpha, \delta) v dv  d\alpha d\delta, \\
E_{\rm tot }&=& \int_{v_{\rm min}}^{v_{\rm max}} \int_{\alpha,\delta}  \frac{1}{2}M(v, \alpha, \delta) v^2 dv  d\alpha d\delta.
\end{eqnarray}

%In section \ref{results} we discuss the validity of these approximations. 

\section{Results}\label{results}

\subsection{Mass versus Time}

We observed the simulation at seven different times distributed from 0.2 Myr to 0.5 Myr, which spans the main accretion period of the forming protostar. In the fiducial case the outflow is inclined  $20-30\degree$ with respect to the line of sight. We define the outflow as any emitting gas with velocity $|v|>1 \, \kms$ along the line of sight. This is consistent with observational determinations of the outflow mass from molecular emission which typically adopt a velocity cutoff of $1-5 \, \kms$  \citep[e.g.,][]{hatchell07,arce10,vandermarel13,dunham14}. This is the minimum velocity for which the outflow can be reliably separated from the embedding turbulent cloud \citep{arce02}. In the simulation,  the turbulent rms core velocity is $\sim 0.5\, \kms$ at the time of the initial outflow launch.

Under this outflow definition, the derived mass initially increases as material is swept up and accelerated, reaches a peak around $0.25$ Myr, and then declines when most of the core has been accreted.  The decline is due to a combination of diminishing accretion, and hence outflow launching, and less material available for entrainment. We stress that this definition, which follows observational conventions, does not directly measure the full protostellar wind information, but rather the emission from the CO in the cloud entrained by the protostellar wind.
%component that is projected along the line-of-sight with emission from entrained CO.  

Figure \ref{velcut} shows the $^{13}$CO outflow mass for velocities $|v_{\rm}|\ge 1,1.5, 2\, \kms$. All exhibit similar trends of mass versus time. At early times the derived mass overestimates the underlying gas mass. This is mainly due to the assumption of LTE, which overestimated the line emission. We discuss this further in the following section.

%Figure \ref{fig3} shows that our observations improve as we increase the cutoff, though we found that we still overestimate the mass slightly.

\begin{figure}[h!]
\begin{center}
\includegraphics[width=1.0\columnwidth]{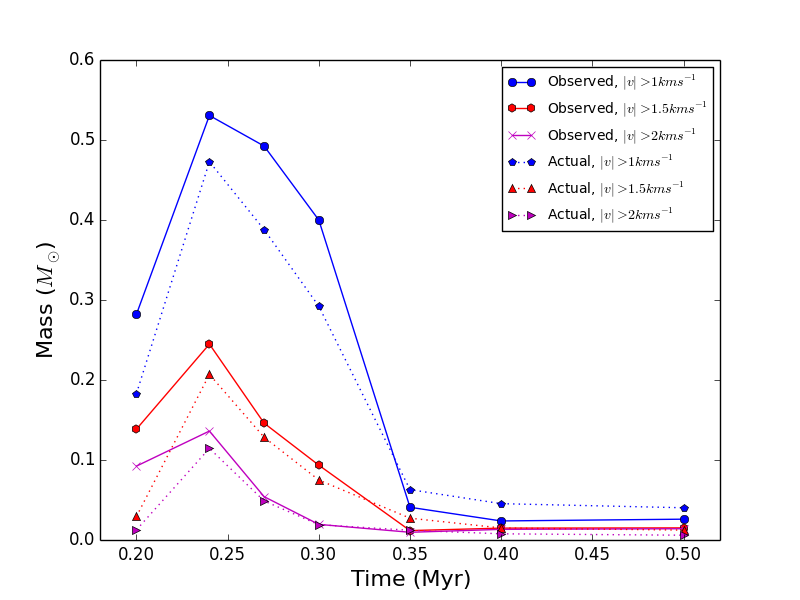}
\caption{Observed (solid) and simulated (dotted) outflow mass versus time with various cutoff velocities.  The observed mass was derived using the fiducial full ALMA configuration (R26 in Table \ref{tab_runs}).   The actual mass of the simulation along the same sightline for each cutoff is indicated by the dotted lines.
\label{velcut}}
\end{center}
\end{figure}

Figure \ref{mass_percent} shows the fraction of the total mass estimated from $^{13}$CO as a function of velocity. Approximately $20\%$ of the mass has a line-of-sight (los)  velocity above $1\,\kms$ and 5\% has a velocity above $2 \kms$. This fraction is somewhat sensitive to the observing configuration. Configurations that resolve the large-scale structure well (e.g., R26) recover more of the emission at lower velocities.  
 All observations miss the very low velocity gas, which has the highest optical depth, but most configurations detect similar amounts of higher velocity gas.

\begin{figure}[h!]
\begin{center}
\includegraphics[width=1.0\columnwidth]{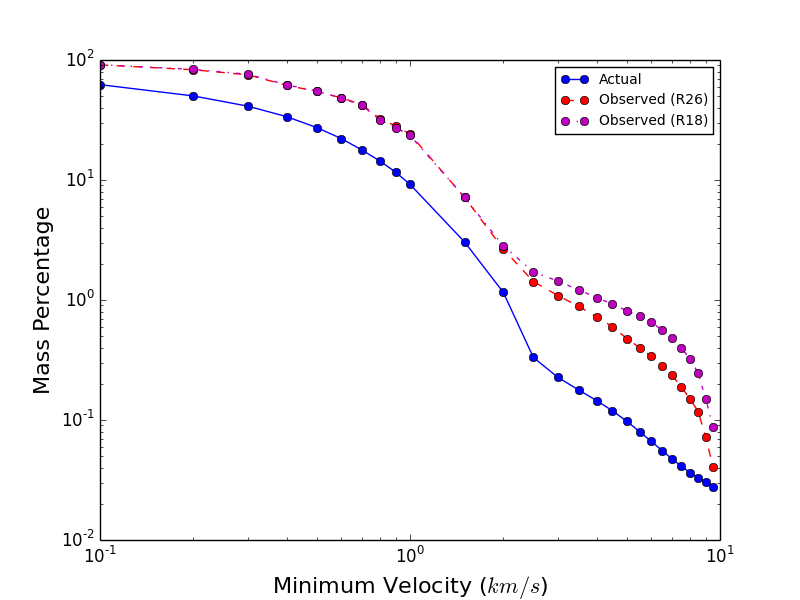}%{figures/Msses2/Msses2.png}
\caption{Percentage of all mass in the domain with velocity greater than various minima for two synthetic observations and the simulation in $^{13}$CO. Observations were made at $t=0.27$ Myr with the R26 and R18 configurations.
%\caption{Cumulative mass percentage with increasing velocity for two synthetic observations and the simulation in $^{13}$CO. Observations were made at $t=0.27$ Myr with the R26 and R18 configurations.}
 \label{mass_percent}}
\end{center}
\end{figure}

\subsection{Excitation Temperature}\label{results_texcit}

Observational derivations of the outflow mass typically adopt a single excitation temperature. Under the assumption of LTE, the excitation temperature is identical to the gas temperature. In equation \ref{eqn:massfromtemp}, we adopt a fiducial value of $T_{\rm ex}=$11 K. However, the gas has a range of underlying temperatures and is not guaranteed to be in LTE.  Figure \ref{texcit} illustrates the impact of the assumed excitation temperature on the mass estimate. The higher the assumed excitation temperature, the higher the estimated mass, so errors in the assumed temperature have a large impact on the accuracy of the mass estimate.
%If we adopt the mass-averaged gas temperature for each time (see Table \ref{tab_texcit}), we derive mean temperatures of $\sim$20 of 11-17 K, which are similar to the fiducial one. 
As shown in Table \ref{tab_texcit}, we derive mean and median values of $\sim20$ K and $10-11$ K, respectively, for the overall gas temperature. The mean outflow temperature is significantly higher than the median, which indicates that a minority of warm cells are dominating the average. The very low median outflow temperature underscores the extent to which the outflow entrains cold core gas.  The median and mean outflow temperature are only similar at the last two times when nearly all the core gas has been accreted or expelled, leaving little cold gas for entrainment. Based on our simulation comparison between the observed and actual mass, equation \ref{eqn:massfromtemp} gives a more accurate mass assessment when a value close to the median temperature is adopted and systematically overestimate the actual outflow mass for larger values. 

%However, since the total mass is dominated by the core gas, which is initialized to be 10K, these average values reflect the temperature of the core rather than that of the higher velocity outflow.  Computing the temperature average only for gas with velocity $|v|>1\, \kms$, we obtain average outflow temperatures of up to 50 K. In combination with equation \ref{eqn:massfromtemp}, these more realistic values systematically overestimate the actual outflow mass.

The origin of the disagreement is the assumption of LTE. Using the ratio of the level populations computed by the {\sc RADMC} LVG radiative transfer calculation, we can compute the effective LVG excitation temperature:
\begin{equation}
\frac{n_u}{n_l} = \frac{g_u}{g_l}e^{-h \nu/(k T_{\rm LVG})},
\end{equation}
where $n_u$, $n_l$ are the upper and lower level populations ($u=1$, $l=0$), $g_u$, $g_l$ are the statistical weights of these levels ($g_1/g_0=3$), $\nu$ is the transition frequency ($\nu=\nu_{^{13}\rm CO(1-0)}$), and $T_{\rm LVG}$ is the effective excitation temperature. In LTE,  this temperature is equal to the gas temperature.
Figure \ref{tcomp} shows that the effective excitation temperature is systematically lower than the actual gas temperature, and thus the gas is not in LTE. Much of this difference is due to the outflow, while the slower (core) gas is closer to LTE. Thus, adopting an excitation temperature that is comparable to or larger than the gas temperature causes equation \ref{eqn:massfromtemp} to over-estimate the outflow mass. As a secondary check, we repeat the radiative transfer calculation under the assumption that the level populations are in LTE. We find that outflow mass estimated from equation \ref{eqn:massfromtemp} for a 7200s observation at 0.24 Myr is $\sim$10\% lower than when the radiative transfer is performed assuming LVG, and consequently, nearly identical to the actual outflow mass. 

In a comparison between simulated outflows and CO synthetic observations \citet{peters14} also found that estimated physical quantities exceeded the actual values, often by factors of 2-3. They do not explore the origin of this discrepancy,  however,  they also adopt an equation similar to \ref{eqn:massfromtemp}, which also assumes LTE. Consequently, it is likely that the underlying level populations in their calculations are likewise not in LTE. \footnote{They adopt a much warmer excitation temperature than what we adopt here (50 K). However, their simulated gas is initially two times warmer and the mean total gas temperature is comparable to or exceeds the assumed excitation temperature in two of their three cases.} 

%We attribute this to errors in assuming a single temperature of excitation. 
%We examined the simulation data and found that there was significant temperature variation in the data cube and slight temperature variation over time. We followed the same protocol as observers and assumed a single temperature of excitation for each image. However we used a different temperature for each time. These temperatures are shown in Table \ref{tab_texcit}.

\begin{deluxetable}{c  |  c c c c} 
\tablecolumns{5}
\tablecaption{Average gas temperature \label{tab_texcit}}
\tablehead{ \colhead{ $t_{\rm run}$(Myr)\tablenotemark{a}} & 
\colhead{ $ \bar T_{\rm tot}$(K)} & 
\colhead{ $ \bar T_{ |v|\ge1}$(K)} & 
\colhead{Median $T$(K)} & 
\colhead{Median $T_{ |v|\ge1}$(K)} } \\
\startdata
0.20  & 28.22   & 69.56 & 11.25 & 10.30  \\%54077826 \\
0.24  & 18.68 & 41.33 & 10.06 & 10.07 \\%79129249 \\
0.27  & 17.80 & 28.03 & 10.08 & 10.15 \\%28610325 \\
0.30  & 17.08  & 30.15 & 10.08 & 10.13 \\%62020028 \\
0.35  & 19.46  & 68.07 & 10.10 & 11.70 \\%17959508 \\
0.40  & 19.62 & 62.57 &  10.22 & 48.96 \\%38009796 \\
0.50  & 22.84 & 69.90  & 11.52 & 73.86 %22632105 \\
\enddata
\tablenotetext{a}{Simulation time measured from the start of the calculation and mass-weighted average gas temperature and median temperature defined over all velocities and over only high velocities. The average temperature is computed as $\bar T = \frac{ \Sigma \rho_{i} T_{i}  }{\Sigma \rho_{i}}$, where the sum is only taken over gas with $T<300$ K or gas with $T<300$ K  and $|v| > 1\kms$.  }
\end{deluxetable}

\begin{figure}[h!]
\begin{center}
\includegraphics[width=1\columnwidth]{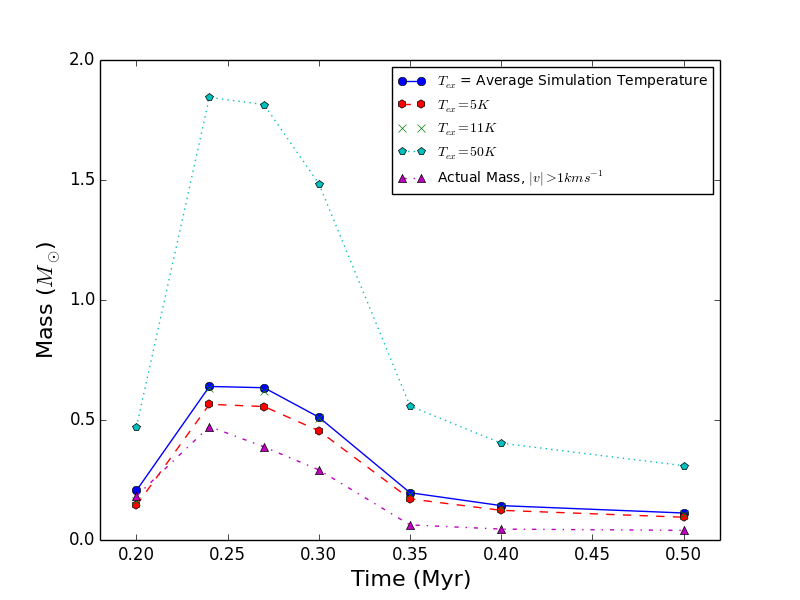}
\caption{Mass with $|v| > 1$ versus time calculated from $^{13}$CO simulated emission for various values T$_{ex}$. The actual simulated mass along the same sightline for $|v| > 1$ is indicated by the dot-dashed line (purple triangles). \label{texcit}}
\end{center}
\end{figure}

\begin{figure}[h!]
\begin{center}
\includegraphics[width=1\columnwidth]{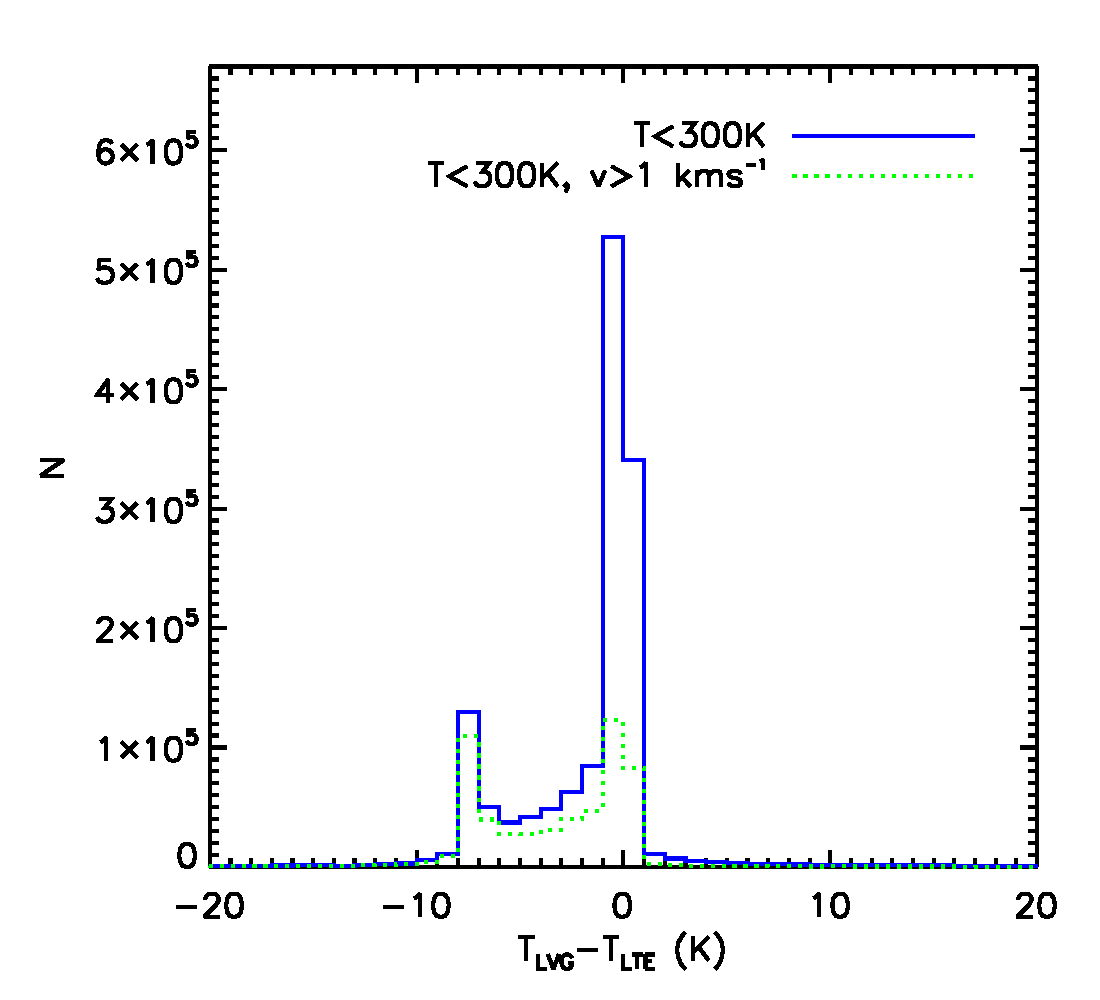}
\caption{ Temperature difference between the effective excitation temperature, $T_{\rm LVG}$ (solid blue), and the actual gas temperature, $T_{\rm LTE}$ (dotted, green), at time 0.24 Myr for cells with temperatures less than 300 K (solid, blue) and cells with temperature less than 300 K and line-of-sight velocities greater than 1$\kms$ (dotted, green). $N$ is the number of cells in each bin.  \label{tcomp}}
\end{center}
\end{figure}

\subsection{Integration Time}

Figure \ref{fig5} illustrates the effect of different total integration times on the estimated mass. As expected, longer observations were able to recover more emission. The difference between a  5 minute and 2 hour observation could be up to a factor of 2 in mass. In both cases, the structure of the outflow cavity is visible, but more of the larger scale structure is resolved out in the shorter observation.  At late times, the 5 minute observation overestimates the mass  due to the low signal-to-noise, which creates false signal  (see Figure \ref{figoutmap}.) In contrast, the 1 and 2 hour integration times have very good signal-to-noise and recover much of the detailed outflow structure.

At the earliest output time, there was little difference in the  mass estimated by the different integration times  since the gas is very optically thick. Even at later times, the total mass is significantly underestimated, since the emission is optically thick in the central region for the low velocity gas at all times.  Consequently, at best half the total mass can be recovered from $^{13}$CO(1-0) even with long observing times.  Otherwise, the gas with $|v|\ge 1\kms$, which is essentially the outflow, is sufficiently optically thin in $^{13}$CO that a 1 hour full ALMA observation is able to recover nearly all of the emission.

\begin{figure}[h!]
\begin{center}
\includegraphics[width=1\columnwidth]{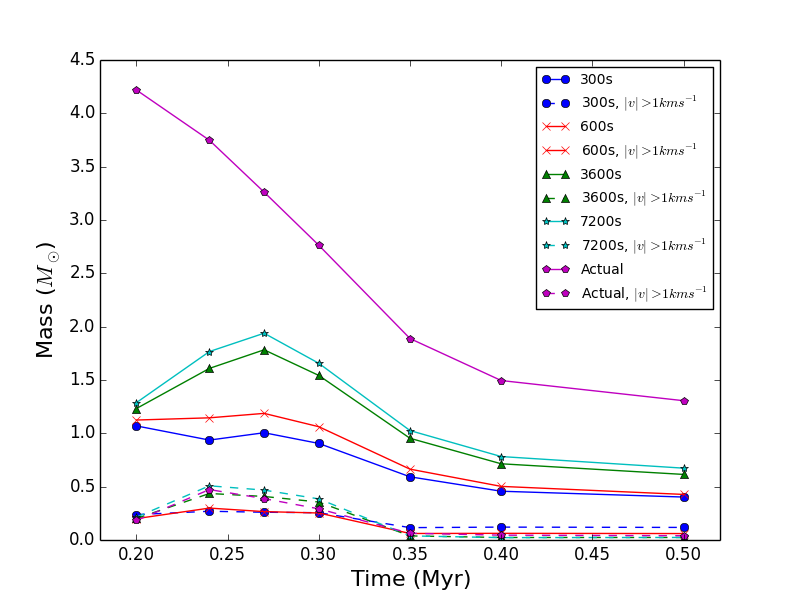}
\includegraphics[width=1.0\columnwidth]{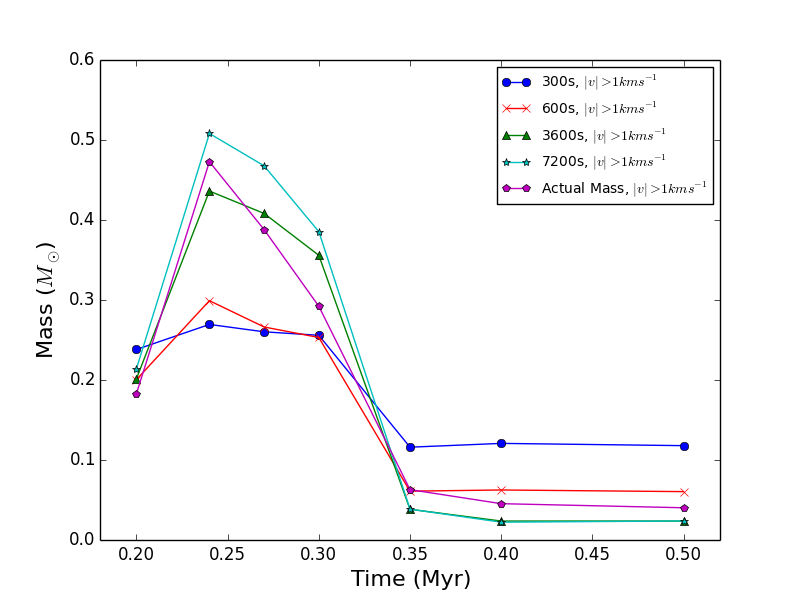}
\caption{Top: (a) Mass versus time derived from $^{13}$CO observed with various integration times. The solid lines show all gas while the dashed lines indicate the mass with $|v|\ge 1 \kms$. Bottom: (b) Mass versus time derived from $^{13}$CO observed with various integration times for gas with $|v|\ge 1 \kms$. The observations are performed with the full ALMA configuration and 10 second pointings (R23-R26). The actual simulation mass is indicated by the purple solid line with star symbols, and the actual simulation mass with $|v|\ge 1 \kms$ is indicated by the purple dashed line with star symbols. \label{fig5}}
\end{center}
\end{figure}

\begin{figure}[h!]
\begin{center}
\includegraphics[width=1.0\columnwidth]{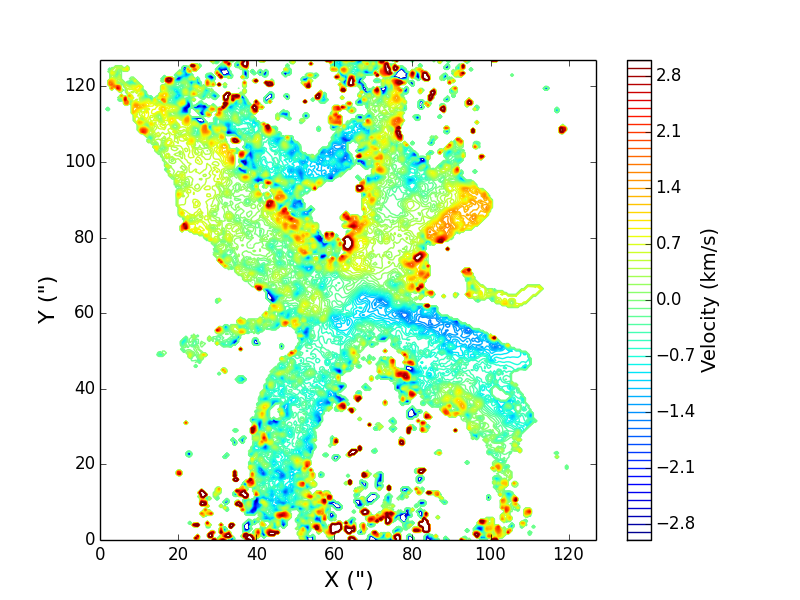}
\includegraphics[width=1.0\columnwidth]{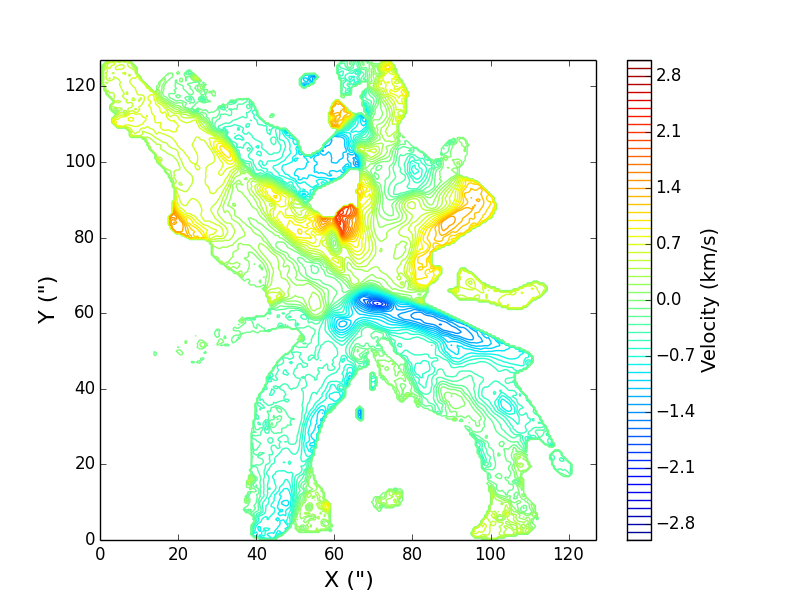}
\caption{Top: (a) $^{13}$CO emission contour plot, colored by velocity for $t = 0.27$Myr. The synthetic observation uses the full ALMA configuration 3, 300s integration time, and 10 second pointings (R24). 
  Bottom: (b) $^{13}$CO emission contour plot, colored by velocity for $t = 0.27$Myr. The synthetic observation uses the full ALMA configuration 3, 7200s integration time, and 10 second pointings (R26). 
 Velocity is calculated as the mass weighted average at each position. The first contour is at $-3 \kms$ with a step of $0.1 \kms$.
 \label{figoutmap}}
\end{center}
\end{figure}

\subsection{ALMA Configuration}

The choice of ALMA configuration has a significant effect on the quality of the observation.
Table \ref{tab_config} lists the four ALMA configurations in three different observing cycles that we compare. 

%replace area with minimum baseline?
\begin{deluxetable*}{c | c c c c c c}
\tablecolumns{7}
\tablecaption{ALMA Configurations \label{tab_config}}
\tablehead{ \colhead{ Configuration\tablenotemark{a}} & \colhead{ $N_{\rm antenna}$} & \colhead{ $\bar d_{\rm sep}$(m)} & \colhead{ $d_{\rm max}$(m)} & \colhead{ $d_{\rm min}$(m)} & \colhead{ $\theta$(")} &  \colhead{ Area(m$^2$)}} 
  \startdata
  Cycle 0 Compact & 16 & 56.9 & 126 & 18.3 & 5.23 &  7240 \\
  Cycle 1.1       & 32 & 71.4 & 166 & 15.1 & 3.99 & 14500 \\
  Cycle 1.3       & 32 & 164  & 443 & 21.3 & 1.49 & 14500 \\
  Full 3          & 50 & 93.2 & 260 & 15.0 & 2.54 & 22600
\enddata
\tablenotetext{a}{Antenna configuration, total number of antennas, average dish separation, maximum baseline, minimum baseline, effective beam resolution, and total area of the various ALMA configurations. To get these, We use the antenna configuration files provided with CASA.}
 \end{deluxetable*}
% Configuration files in Mac here:
%/Applications/CASA.app/Contents/data/alma/simmos/ 

An outflow spanning a few $0.1$pc at a distance of 450 pc is well-mapped by a beam of a several arcseconds.  Figure \ref{cycle1v3}a shows the mass estimated from  the Cycle 1.1 and 1.3 configurations. Configuration 1.3 (R17) with its longer baselines and smaller beam was significantly less effective than the Cycle 1.1 (R14) observation. As illustrated by Figure \ref{mapcycle1v3}, although the smaller beam can recover more small scale detail it resolves out most of the larger scale emission, including much of the outflow cavity structure. Consequently, the signal-to-noise is significantly lower.

\begin{figure}[h!]
\begin{center}
\includegraphics[width=1.0\columnwidth]{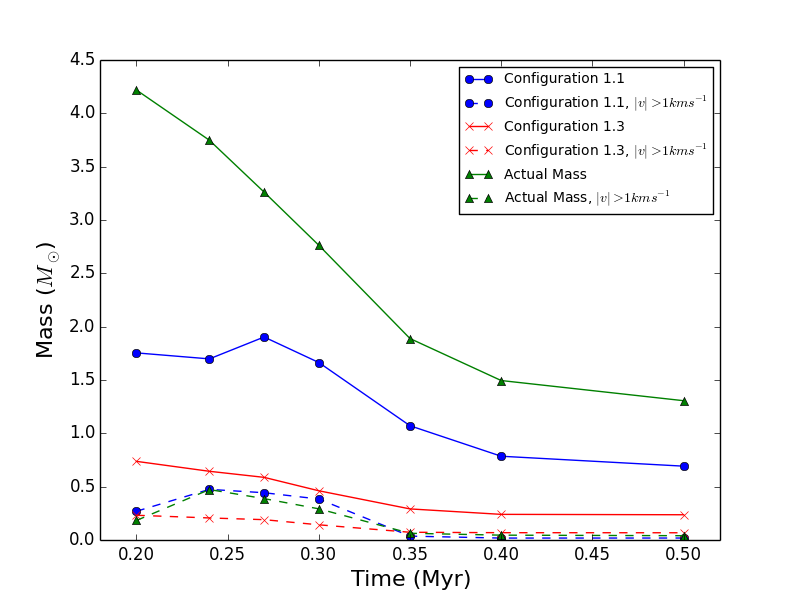}
\includegraphics[width=1.0\columnwidth]{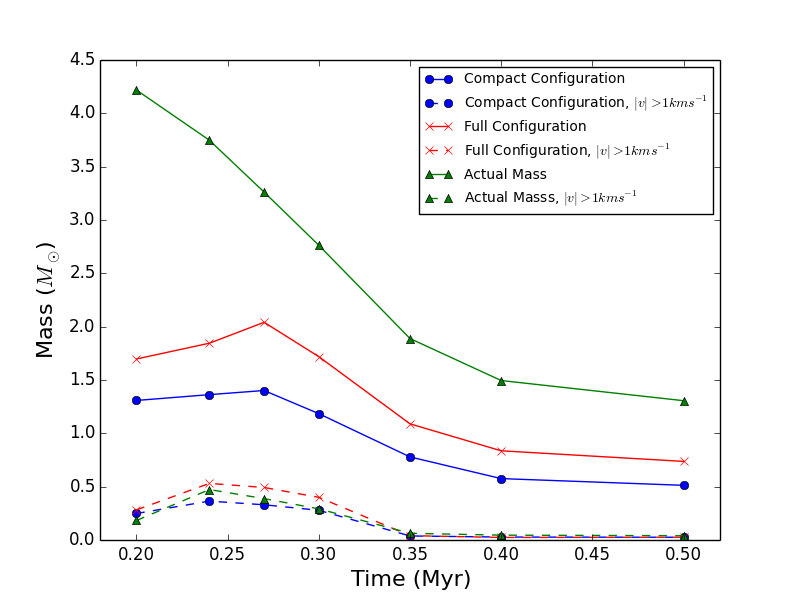}
\caption{Top: (a) Mass versus time for $^{13}$CO observations performed with the Cycle 1.1 (blue circles) and Cycle 1.3 (red crosses) configurations and an integration time of 7200s with 30s pointings. The actual mass is indicated by the green triangles.  Bottom: (b) Mass versus time of synthetic ALMA observations. The observations were performed with the Full Array configuration 3 and the Cycle 0 Compact configuration with an integration time of 7200s and 10s pointings. The actual mass is indicated by the green triangles. \label{cycle1v3}}
\end{center}
\end{figure}

\begin{figure}[h!]
\begin{center}
\includegraphics[width=1.0\columnwidth]{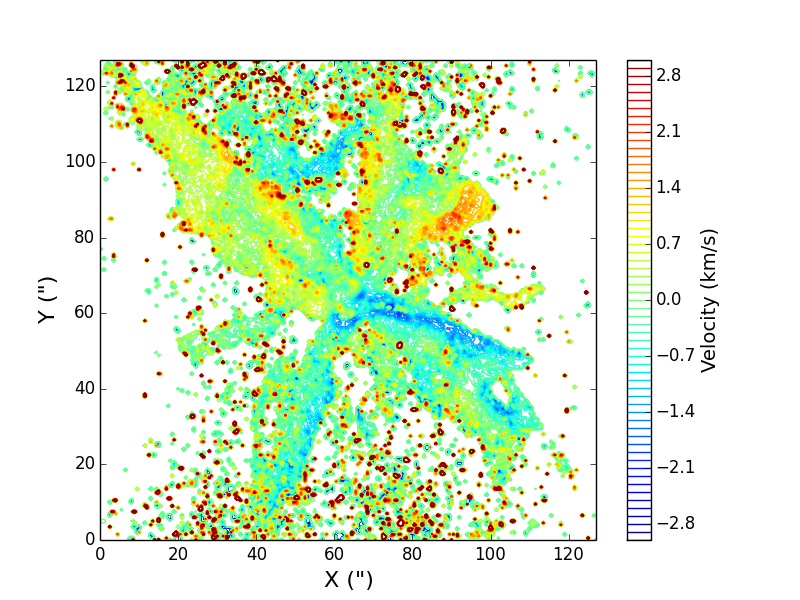}
\includegraphics[width=1.0\columnwidth]{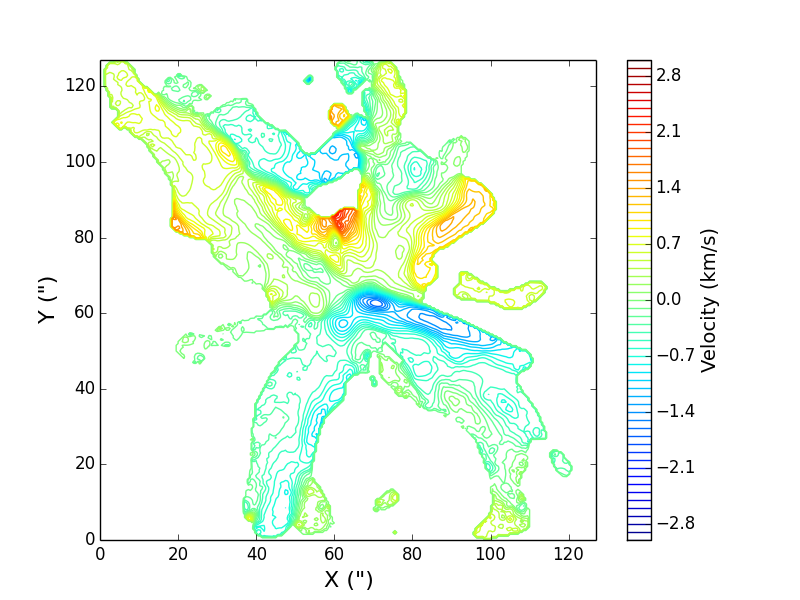}

\caption{Top: (a) $^{13}$CO emission contour plot, colored by velocity for $t = 0.27$Myr. The synthetic observation is performed with the 1.3 ALMA configuration, over 7200s with 30 second pointings (R17).
  Bottom: (b) $^{13}$CO emission contour plot, colored by velocity for $t = 0.27$Myr. The synthetic observation is performed with the 1.1 ALMA configuration over 7200s with 30 second pointings (R15).
  Velocity is calculated as the mass weighted average at each position. The first contour is at $-3 \kms$ with a step of $0.1 \kms$.
\label{mapcycle1v3}}
\end{center}
\end{figure}

As expected, increasing the number of antennas further increases the emission recovery. For example, despite having a smaller effective beam, full ALMA in configuration 3 produces a very similar map to the Cycle 1.1 configuration map (compare Figure \ref{mapcycle1v3}b with Figure \ref{figoutmap}b ).
As illustrated by Figure \ref{cycle1v3}b, full ALMA performs significantly better with higher signal-to-noise than Cycle 0 for the same integration and pointing times, and a factor of $\sim$2 more mass is recovered.

\subsection{$^{12}$CO(1-0) }

$^{12}$CO is often used to estimate outflow mass. While $^{12}$CO(1-0) is generally very optically thick in dense cores, it is less optically thick over the higher velocity outflow gas (e.g, $|v|\gtrsim2\kms$), and thus the assumption of optical thinness and LTE is often adopted  \citep[e.g.][]{lada96,vandermarel13}. 
This assumption is more likely to be valid for low-resolution observations, in which the unresolved emission is spread over a larger area by the beam. However, at the high-resolution of ALMA 
we find that even the gas above a couple \kms  is optically thick in $^{12}$CO(1-0). Consequently, without an opacity correction the $^{12}$CO severely underestimates the outflow mass. Figure \ref{asdf} shows the mass versus time for observations in $^{12}$CO and $^{13}$CO if we assume that both are optically thin. These results clearly demonstrate that $^{12}$CO is optically thick, even for gas with $|v|  \ge1$.

\begin{figure}[h!]
\begin{center}
\includegraphics[width=1.0\columnwidth]{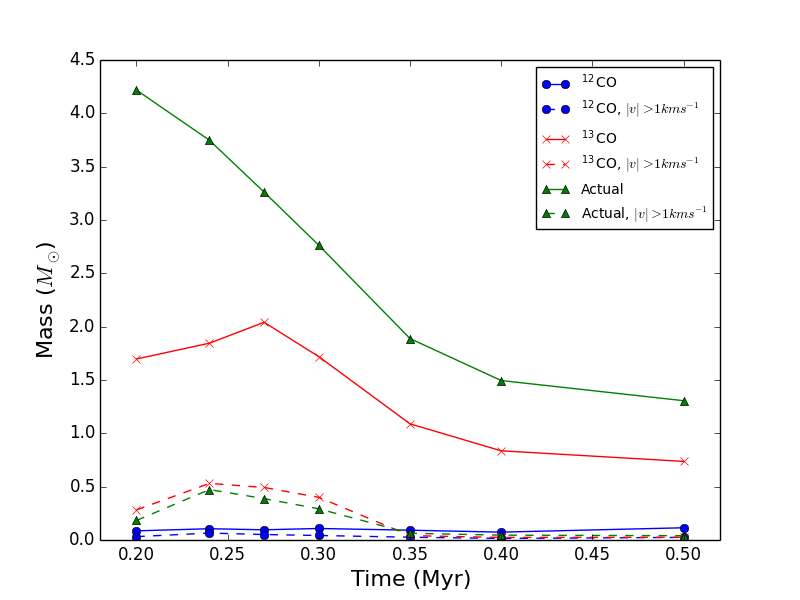}
\caption{Mass versus time for synthetic $^{13}$CO and $^{12}$CO observations (R26, R24). The synthetic observations are performed with full ALMA configuration 3, 10 second pointings and 7200s total observation time. The green triangles indicate the actual simulation mass. \label{asdf}}
\end{center}
\end{figure}

However, it is possible to use the optically thin $^{13}$CO emission to estimate the optical depth and then correct the $^{12}$CO mass estimate. Given the brightness temperatures for each line, the relation between the brightness temperatures and the optical depth, $\tau_{12}$, can be expressed as \citep{dunham14}:
\begin{equation} \label{eqn:mass} 
\frac{1 - e^{-\tau_{12}}}{\tau_{12}} = \frac{ T_{\rm mb 12}}{T_{\rm mb 13}} \frac{[ ^{13}{\rm CO}]}{[^{12}{\rm CO}]}.\,
\end{equation}
where $\frac{[ ^{13}{\rm CO}]}{[^{12}{\rm CO}]}$ is the isotopic ratio.
We can then correct the $^{12}$CO data by estimating the optical depth of each voxel and multiplying the emission in this velocity channel by the appropriate factor. Figure \ref{corrCOmass} shows the estimated mass after this correction is applied.

\begin{figure}[h!]
\begin{center}
\includegraphics[width=1.0\columnwidth]{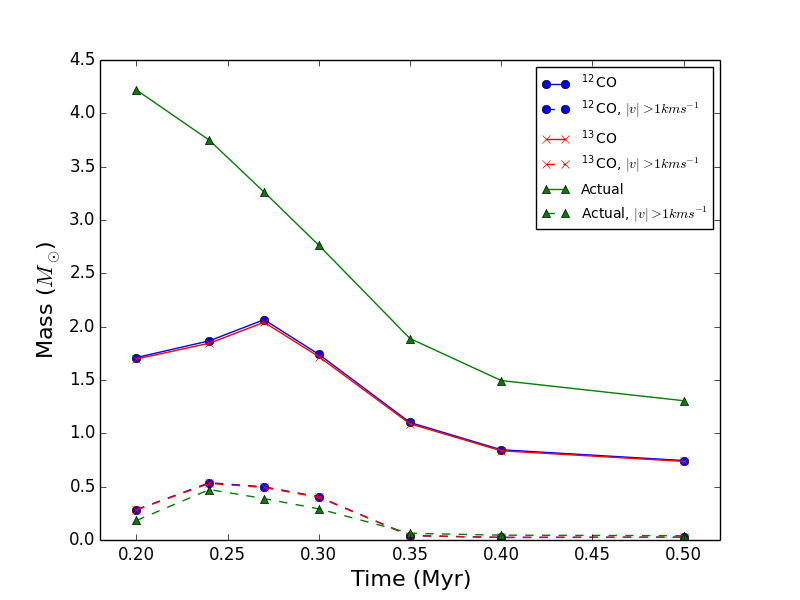}
\caption{Mass versus time for synthetic $^{13}$CO and $^{12}$CO observations  (R26, R24), where the $^{12}$CO emission is corrected using equation \ref{eqn:mass}. The synthetic observations are performed with full ALMA configuration 3, 10 second pointings and 7200s total observation time. The green triangles indicate the actual simulation mass. \label{corrCOmass}}
\end{center}
\end{figure}

Figure \ref{corrCO} shows the pre- and post-opacity corrected emission maps for $^{12}$CO. Without the correction much of the outflow structure is missing simply because it has a low signal-to-noise. Applying the correction factor systematically boosts the emission from the relatively dense outflow cavity walls,  and the resulting map looks much more similar to that of $^{13}$CO map (e.g. bottom panel of Figure \ref{mapcycle1v3}). Consequently, the $^{12}$CO is only useful for determining the velocity and structure of the outflow (at high resolution) when it is possible to measure the optical depth and apply a correction factor. The $^{13}$CO observations, which in most cases do not require a correction, are significantly more accurate. 
%Single-dish observations of $^{12}$CO(1-0), which smear the emission over a much larger area, may be more optically thin at these velocities. 
%{\bf I am not sure about this statement. I do not think it is correct. I would not include it unless you actually tested it to see if was true that for larger beams the 12CO is less optically thin}.

\begin{figure}[h!]
\begin{center}
\includegraphics[width=1.0\columnwidth]{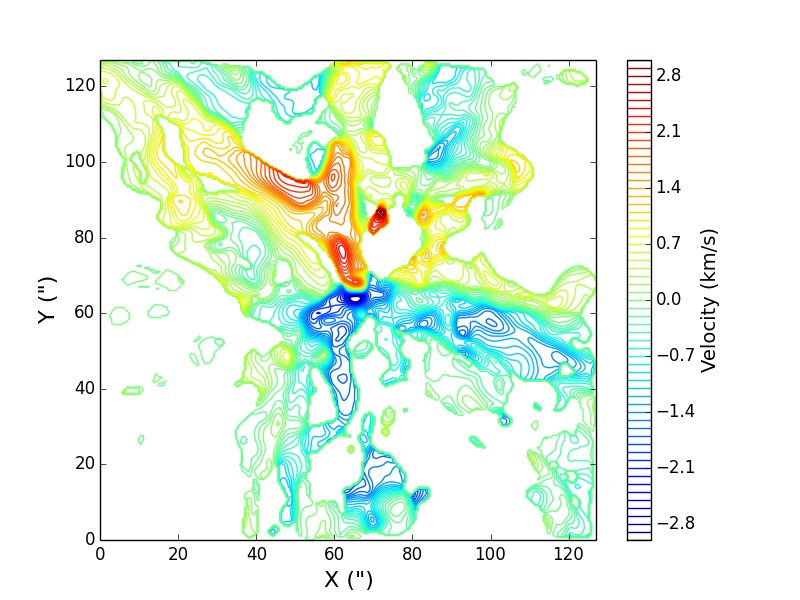}
\includegraphics[width=1.0\columnwidth]{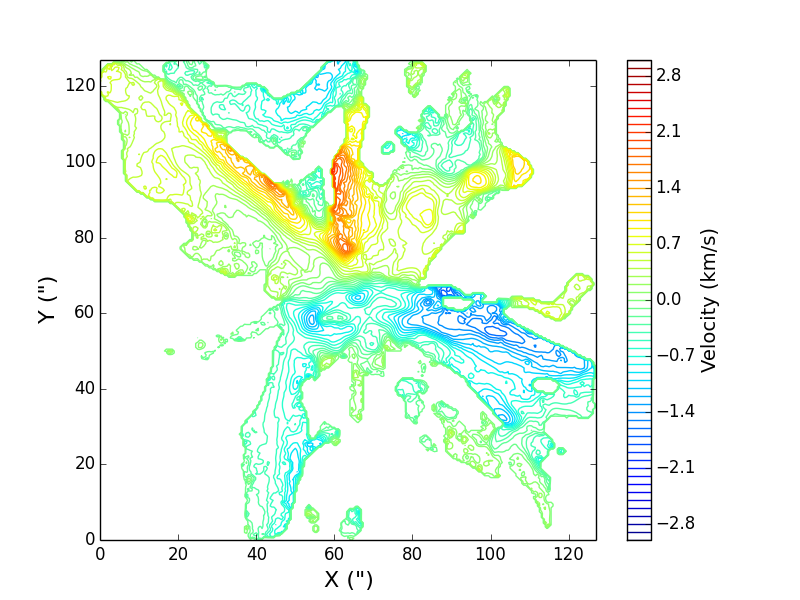}
\caption{Top: (a) Uncorrected $^{12}$CO emission contour plot, colored by velocity for $t = 0.3$Myr. 
      Bottom: (b) Corrected (using equation \ref{eqn:mass}) $^{12}$CO emission contour plot, colored by velocity for $t = 0.3$Myr.
The observations use the full ALMA configuration 3, 7200s integration time, and 10 second pointings (R4).
Velocity is calculated as the mass weighted average at each position. The first contour is at $-3 \kms$ with a step of $0.1 \kms$.
\label{corrCO}}
\end{center}
\end{figure}

\subsection{Viewing Angle}

The accuracy of the mass recovery depends partially on the angle of the outflow axis along the line-of-sight.  Here, we compare two different viewing angles. The fiducial case is inclined $20-30 \degree$ towards the line-of-sight and the second is inclined by $35-45\degree$ (the outflow axis is not fixed in the calculation and it varies by $\sim$10\degree).  

\begin{figure}[h!]
\begin{center}
\includegraphics[width=1.0\columnwidth]{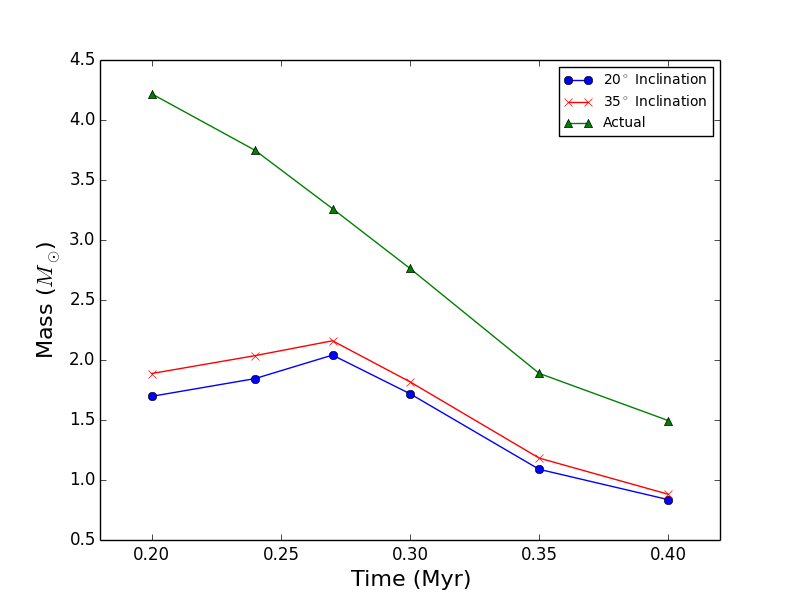}
\caption{Mass versus time for synthetic $^{13}$CO observations for two views for inclinations differing by $\sim15-20\degree$. The synthetic observations use the full ALMA configuration 3, 7200s total observation time, and 10 second pointings.  \label{2views}}
\end{center}
\end{figure}

\begin{figure}[h!]
\begin{center}
\includegraphics[width=1.0\columnwidth]{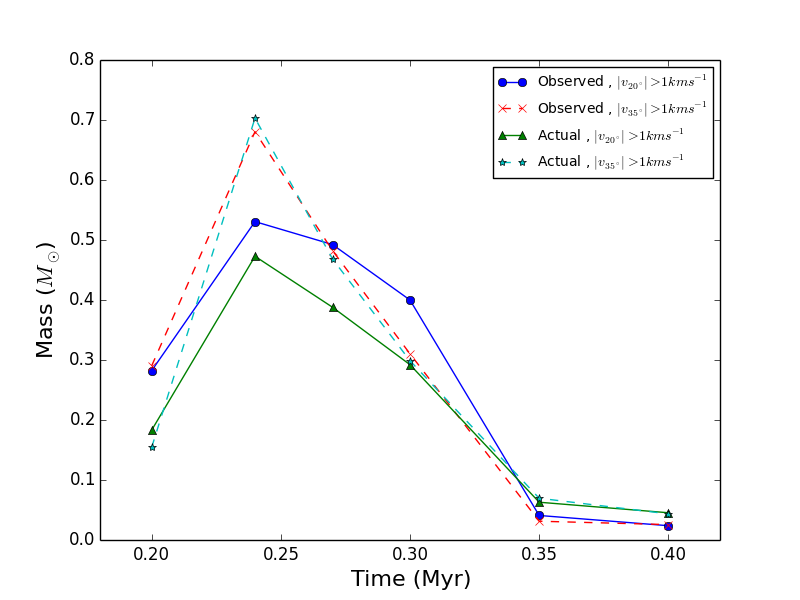}
\caption{Mass $|v| >$ \kms versus time for synthetic $^{13}$CO observations for two outflow inclinations. The synthetic observations are performed with the full ALMA configuration 3, 10 second pointings and 7200s total observation time.}
\end{center}
\end{figure}

While we expect slightly different results for the mass for different views, Figure \ref{2views} shows that these particular sight-lines yield similar estimates. As long as the molecular emission is optically thin, as in the case of $^{13}$CO for $|v| \ge1$ \kms, the estimated mass should be similar for outflow inclinations from $15-75 \degree$. At extreme inclinations the mass is far harder to estimate. Outflows viewed perpendicular to the axis ($0\degree$) are difficult to separate from the ambient material because only a small component of the velocity is projected along the line-of-sight. Outflows viewed down the axis ($90\degree$) are easier to identify since the highest component of the velocity is along the viewing direction. However, at this inclination the two outflow lobes are projected on top of one another complicating the measurement of the gas morphology.

\begin{deluxetable*}{c | c  c  c  c  c  c  c  c} 
\tablecolumns{9}
 \tablecaption{Derived Physical Parameters\tablenotemark{a} \label{tab_prop}}
\tablehead{ \colhead{} &  
\colhead{$M_{\rm sim, tot}$} &  
\colhead{$M_{\rm sim, |v| > 1}$} & 
 \colhead{$M_{^{13}{\rm CO, tot}}$} &  
 \colhead{$M_{^{13} {\rm CO},|v| > 1}$} &  
 \colhead{$M_{^{12}{\rm CO,tot}}$} & 
  \colhead{$M_{^{12}{\rm CO, |v|>1}}$} &  
  \colhead{$M_{^{12} {\rm CO, tot, corr}}$} &  
  \colhead{$M_{^{12}{\rm CO},|v|>1, {\rm corr}}$}} \\
\startdata
 $M$ ($M_{\odot}$) & 3.22 & 0.39 & 2.04 & 0.49 & 0.10 & 0.051 & 2.07 & 0.50 \\
 $P$ ($M_{\odot}$ \kms) & 1.57 & 0.59 & 1.38 & 0.79 & 0.12 & 0.10 & 1.39 & 0.80 \\
 $E$ ($10^{42}$ erg) & 18.75 & 13.45 & 19.75 & 16.47 & 2.95 & 2.84 & 19.98 & 16.67 
\enddata
\tablenotetext{a}{Mass, momentum, and energy for different mass definitions.  All emission estimates are calculated with full ALMA configuration 3 (R26) with an integration time of 7200s and a pointing time of 10s for $t_{\rm run}$=0.27Myr.}
\end{deluxetable*}

% Table with all the mass info
\begin{deluxetable*}{c | c c c c c c c c} 
\tablecolumns{9}
 \tablecaption{Derived Mass\tablenotemark{a} \label{tab_mass}}
\tablehead{ \colhead{ $t_{\rm run}$ (Myr)} & 
\colhead{$M_{\rm tot}$} &  
\colhead{$M_{|v|>1}$} &   
\colhead{$M_{^{13}{\rm CO}}$} &  
\colhead{$M_{^{13}{\rm CO}, |v|>1}$}  &  
\colhead{$M_{^{12}{\rm CO}}$}  &  
\colhead{$M_{^{12}{\rm CO}, |v|>1}$} & 
 \colhead{$M_{^{12}{\rm CO, corr}}$}&  
 \colhead{$M_{^{12}{\rm CO},|v|>1, corr}$}} \\
\startdata
0.20 & 4.22 & 0.18 & 1.70 & 0.28 & 0.08 & 0.03 & 1.71 & 0.28 \\ 
0.24 & 3.75 & 0.47 & 1.84 & 0.53 & 0.10 & 0.06 & 1.87 & 0.54 \\ 
0.27 & 3.26 & 0.39 & 2.04 & 0.49 & 0.10 & 0.05 & 2.06 & 0.50 \\ 
0.30 & 2.76 & 0.29 & 1.718 & 0.40 & 0.11 & 0.04 & 1.74 & 0.40 \\ 
0.35 & 1.89 & 0.06 & 1.01 & 0.04 & 0.09 & 0.03 & 1.10 & 0.04 \\ 
0.40 & 1.49 & 0.04 & 0.84 & 0.02 & 0.07 & 0.01 & 0.84 & 0.02 \\ 
0.50 & 1.31 & 0.04 & 0.74 & 0.03 & 0.11 & 0.02 & 0.74 & 0.03 
\enddata
\tablenotetext{a}{Output time, total simulated outflow mass, simulated outflow mass for $|v_z|>1$, total mass estimated from $^{13}$CO,  mass estimated from $^{13}$CO  for $|v_z|>1$, total mass estimated from $^{12}$CO with no opacity correction,  mass estimated from $^{12}$CO  for $|v_z|>1$ with no opacity correction, total mass estimated from $^{12}$CO  with opacity correction,  mass estimated from $^{12}$CO  for $|v_z|>1$ with opacity correction. All emission estimates are calculated with full ALMA configuration 3 (R26) with an integration time of 7200s and a pointing time of 10s for $t_{\rm run}$=0.27Myr. All masses are in units of $\msun$.}
\end{deluxetable*}

% Table for momentum

\begin{deluxetable*}{c | c c c c c c c c} 
\tablecolumns{9}
 \tablecaption{Derived Momentum\tablenotemark{a} \label{tab_momentum}}
\tablehead{ \colhead{$t_{\rm run}$(Myr)} & 
\colhead{$P_{\rm sim, tot}$} & 
\colhead{$P_{sim,|v_z|>1}$} & 
\colhead{$P_{^{13}\rm CO}$} & 
\colhead{$P_{^{13}\rm CO, |v_z| \ge 1}$} & 
\colhead{$P_{^{12}\rm CO}$}  & 
\colhead{$P_{^{12}\rm CO, |v_z| \ge1}$ } & 
\colhead{$P_{^{12}\rm CO}$corr }& 
\colhead{$P_{^{12}\rm CO, |v_z| \ge 1}$,corr }}\\
\startdata    
0.2 & 1.40 & 0.26 & 1.10 & 0.60 & 0.13 & 0.11 & 1.11 & 0.60 \\ 
0.24 & 1.85 & 0.79 & 1.44 & 0.92 & 0.15 & 0.13 & 1.46 & 0.93 \\ 
0.27 & 1.56 & 0.59 & 1.36 & 0.79 & 0.12 & 0.10 & 1.39 & 0.80 \\ 
0.3 & 1.33 & 0.42 & 1.11 & 0.59 & 0.10 & 0.07 & 1.12 & 0.60 \\ 
0.35 & 0.76 & 0.11 & 0.50 & 0.10 & 0.07 & 0.04 & 0.50 & 0.10 \\ 
0.4 & 0.52 & 0.07 & 0.35 & 0.09 & 0.04 & 0.02 & 0.35 & 0.10 \\ 
0.5 & 0.43 & 0.07 & 0.31 & 0.10 & 0.07 & 0.03 & 0.31 & 0.10 
\enddata
%\medskip
\tablenotetext{a}{Output time, total simulated outflow momentum, simulated outflow momentum for $|v_z|>1$, total momentum estimated from $^{13}$CO,  momentum estimated from $^{13}$CO for $|v_z|>1$, total momentum estimated from $^{12}$CO with no opacity correction,  momentum estimated from $^{12}$CO  for $|v_z|>1$ with no opacity correction, total momentum estimated from $^{12}$CO  with opacity correction,  momentum estimated from $^{12}$CO  for $|v_z|>1$ with opacity correction. All emission estimates are calculated with full ALMA configuration 3 (R26) with an integration time of 7200s and a pointing time of 10s for $t_{\rm run}$=0.27Myr. All momenta are in units of $M_{\odot}$kms$^{-1}$.}
\end{deluxetable*}

% Table for energy
 \label{tab_energy}
\begin{deluxetable*}{c | c c c c c c c c}
\tablecolumns{9}
\tablecaption{ Derived Energy\tablenotemark{a}  \label{tab_energy}}
\tablehead{ \colhead{$t_{\rm run}$(Myr)} & 
 \colhead{$E_{\rm sim,tot}$ } & 
 \colhead{$E_{sim, |v_z|>1}$ } & 
 \colhead{$E_{^{13}\rm CO}$} & 
 \colhead{$E_{^{13}\rm CO |v_z| > 1}$ } & 
 \colhead{$E_{^{12}\rm CO}$} & 
 \colhead{$E_{^{12}\rm CO |v_z| > 1}$ } &
 \colhead{$E_{^{12}\rm CO}$,corr} & 
 \colhead{$E_{^{12}\rm CO |v_z| > 1}$,corr }} \\
\startdata
0.2 & 12.43 & 6.98 & 22.16 & 19.58 & 6.16 & 6.06 & 22.34 & 19.73 \\ 
0.24 & 23.12 & 17.52 & 22.91 & 19.92 & 3.19 & 3.09 & 23.19 & 20.16 \\ 
0.27 & 18.75 & 13.44 & 19.74 & 16.47 & 2.95 & 2.84 & 19.98 & 16.66 \\ 
0.3 & 15.65 & 10.53 & 13.40 & 10.48 & 1.47 & 1.32 & 13.56 & 10.61 \\ 
0.35 & 7.15 & 3.85 & 6.435 & 4.31 & 1.10 & 0.96 & 6.51 & 4.36 \\ 
0.4 & 4.20 & 2.08 & 6.67 & 5.52 & 0.48 & 0.35 & 6.74 & 5.58 \\ 
0.5 & 3.68 & 1.99 & 6.84 & 5.89 & 0.70 & 0.51 & 6.91 & 5.96 
\enddata
\tablenotetext{a}{Output time, total simulated outflow energy, simulated outflow energy for $|v_z|>1$, total energy estimated from $^{13}$CO,  energy estimated from $^{13}$CO  for $|v_z|>1$, total energy estimated from $^{12}$CO with no opacity correction,  energy estimated from $^{12}$CO  for $|v_z|>1$ with no opacity correction, total energy estimated from $^{12}$CO with opacity correction,  energy estimated from $^{12}$CO for $|v_z|>1$ with opacity correction. Energy is in units of $10^{42}$ erg. All emission estimates are calculated with full ALMA configuration 3 (R26) with an integration time of 7200s and a pointing time of 10s for $t_{\rm run}$=0.27Myr. }
\end{deluxetable*}

\subsection{Momentum and Energy}

The primary observational uncertainty in determining the line-of-sight momentum and energy relates to how well the mass can be recovered from the emission. In the case of an optically thin (or optical-depth corrected) tracer, we find that ALMA will be able to recover the mass and outflow structure quite well for reasonable observing times and resolutions.  However, estimating the {\rm total} outflow momentum and energy is more fraught for several reasons. First, only one velocity component is directly observable, and the other two must be inferred using projection arguments and estimates of the inclination. Second, a significant fraction of the momentum may reside in lower velocity entrained material, which is both optically thick and difficult to distinguish from the surrounding dense core. If the source is sufficiently isolated, it is possible to model and include this material in the estimate \citep[e.g.][]{arce01b,Offner11,dunham14}, however, accuracy relies on the assumed gas distribution of the parent core. Finally, outflow momentum and energy can reside in ionized or very low-abundance warm gas, which is difficult to detect. In the simulations we analyze, the highest velocity material does not emit in CO, a situation which worsens at late times when there is little residual cold gas available for entrainment. This gas contains a small amount of the outflow mass but a significant fraction of the outflow momentum and energy. Here, we simply compare the line-of-sight momentum and energy, with the caveat that the total amount of momentum and energy accounting for these factors is much higher.

Tables \ref{tab_momentum} and \ref{tab_energy} compare the simulation momentum and energy for a particular sightline with that derived from the observations. Tables \ref{tab_momentum} and \ref{tab_energy} represent best-case ALMA observations with good mass recovery. The total momentum is underestimated by $\sim 30$\% when all velocities are considered, while the total energy is comparable or higher than the simulated value.  This suggests that even applying a model for the core gas could underestimate the outflow momentum if the mass is underestimated. The momentum and energy of the $|v|\ge 1\,\kms$ gas is over-estimated due to the LTE approximation as discussed in \S\ref{results_texcit}. 

Once corrected for opacity, $^{12}$CO observations give very similar momentum and energy estimates to the $^{13}$CO ones. Without the correction, $^{12}$CO observations significantly underestimate both by factors of 2-10.

\section{Conclusions}\label{conclusions}

We performed synthetic ALMA molecular line observations of a numerical simulation of a forming protostar in order to evaluate the accuracy of inferred physical quantities.  To produce the synthetic observations we adopt the distance of source HH46/47, which has recently been observed in CO by ALMA. Given it's distance of 450 pc and molecular extent of a few tenths of a parsec, it is a reasonable representative of isolated, local low-mass protostellar outflows.

Following observational convention we adopted a constant excitation temperature of 11 K and assumed the gas was in local thermodynamic equilibrium (LTE). We demonstrated that higher assumed excitation temperatures can easily increase the apparent mass by factors of 2-3. Inspection of the level populations of the outflow gas (as defined by gas with line-of-sight velocity $|v| \ge 1\, \kms$) indicate that the gas is not in LTE and that the effective excitation temperature is slightly lower than the actual gas temperature. This results in an over-estimate of the outflow mass, momentum and energy.

By comparing with the actual simulated values, we found that the $^{12}$CO(1-0) line is an inaccurate tracer at high resolution since it is optically thick for all times and for velocities greater than 1\,\kms. In contrast, the higher velocity gas emitting in $^{13}$CO(1-0) is optically thin for nearly all times. Consequently, the synthetic $^{13}$CO observations reliably recovered the gas mass for a variety of angular resolutions, viewing angles, and integration times. Shorter pointing times resulted in slightly better results, but the difference was minor compared to the total integration time.  The $^{12}$CO mass estimates were significantly improved by applying an optical depth correction factor computed from the $^{13}$CO(1-0) line.

Although full ALMA has the ability to achieve sub-arcsecond resolution, we found that significant flux was resolved out for effective beam sizes smaller than $\sim2$ arcseconds. Adding more antennas and using longer integration times increased the estimated mass accuracy and structure recovery significantly. One to two hour integration times recovered twice as much flux (and mass) as 5 and 15 minute observations. The best ALMA configuration we tested (full ALMA, configuration 3), was able to produce a detailed and high signal-to-noise map of the outflow with sufficient flux recovery that it {\it overestimated} the outflow mass by $\sim20$\%. This was due to the assumption of LTE, which enhanced the conversion of flux to mass.

The momentum and energy of the $|v|\ge1\,\kms$ gas were comparable to or greater than the simulated values for the best ALMA resolution. This is encouraging for observational assessments of the turbulent energy injection due to outflows. However, corrections to account for outflow orientation and optical depth (for $^{12}$CO) will be necessary to produce high accuracy estimates of the outflow impact. 

In summary, our analysis indicates that ALMA will allow unprecedented study of molecular outflows with high resolution, accuracy and detail. However, additional uncertainties, including the molecular abundance, fraction of low-velocity material, and mass of the non-molecular component, will remain complicating factors even in the era of ALMA.

\acknowledgements  
The authors thank an anonymous referee for constructive comments.
Support for S.S.R.O. was provided by NASA through Hubble Fellowship grant HF-51311.01 
awarded by the Space Telescope Science Institute, which is operated by the Association of
Universities for Research in Astronomy, Inc., for NASA, under contract NAS 5-26555.  H.G.A. acknowledges support from his NSF CAREER award AST-0845619. The simulations were performed on the Yale University Omega machine; this work was supported in part by the facilities and staff of the Yale University Faculty of Arts and Sciences High Performance Computing Center.

%Shorter pointing times resulted in slightly better results, but the difference was minor compared to the total integration time. 
%If the total integration time can be increased by reducing the overhead of shorter pointings, this would almost certainly be beneficial.

\appendix

\begin{figure}[h!]
\begin{center}
\includegraphics[width=0.47\columnwidth]{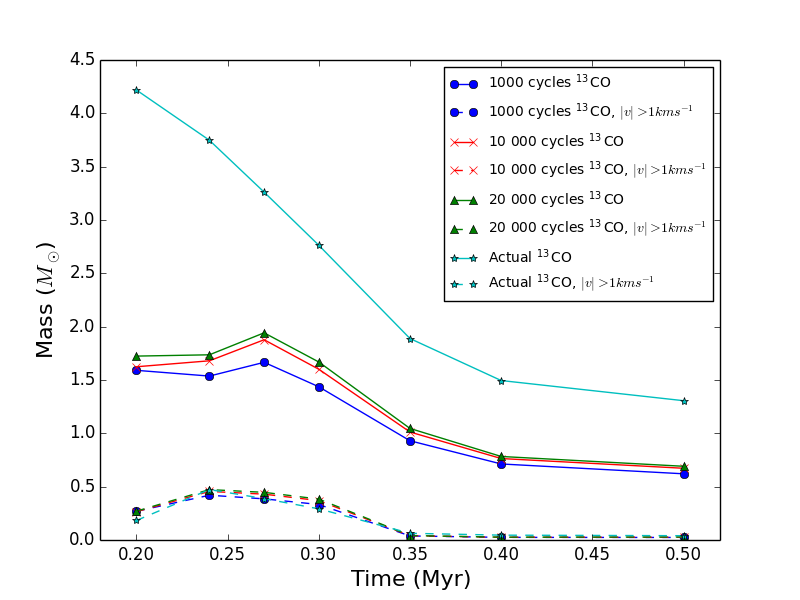}
\includegraphics[width=0.47\columnwidth]{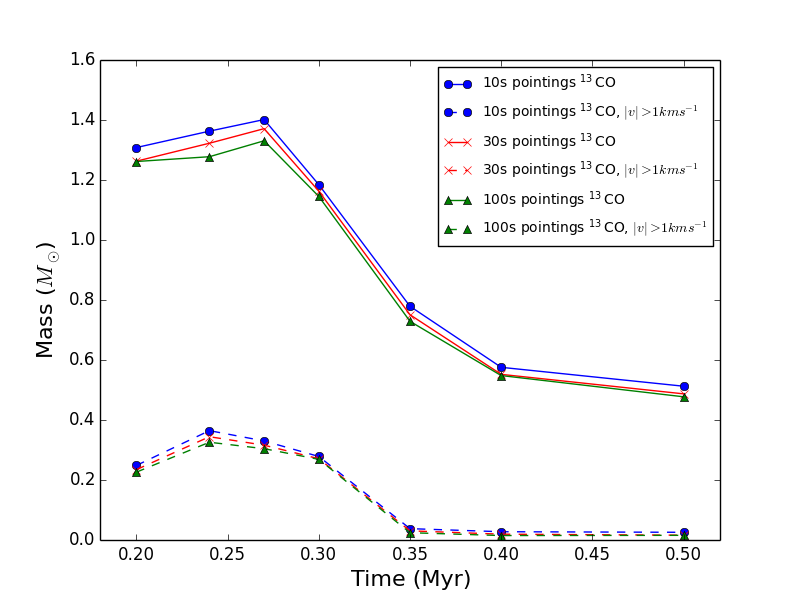}
\caption{Left: (a) Mass versus time for synthetic $^{13}$CO observations with different numbers of cleaning cycles. The synthetic observations use the full ALMA configuration 3, with a 3600s integration time and 10s pointings (R21, R22, R23).  Right: (b) Mass versus time for synthetic $^{13}$CO with various pointing times. The synthetic observations use the Cycle 0 compact ALMA configuration with a 7200s integration time (R18, R19, R20). 
% {\bf Why do you use the Cycle 0 compact ALMA configuration here while for most of the other plots you show the results using the full ALMA configuration 3?}
 \label{figit}}
\end{center}
\end{figure}

There were several additional parameters that we explored but which turned out to have minimal impact. We discuss these parameters here for completeness.  The parameter, $N_{\rm clean}$, sets the limit for the number of cleaning iterations performed by ``simanalyze". We found that changes in the number of iterations had a small affect on the results. Figure \ref{figit}a shows the estimated mass for different values of $N_{\rm clean}$. We adopted $N_{\rm clean}=$10,000 as it seems to provide a balance between good results and efficient use of time and processor cycles.

The parameter $F_{\rm clean}$ sets the threshold for early termination of the cleaning. Our calculations never satisfied the early termination for any  of $0.01-0.1$mJy, so we adopted a default value of 0.1 mJy.

We also considered the effect of different pointing times. The CASA documentation suggests that the pointing time could have a significant effect on the observation, with shorter pointings giving better results. However, we found that the difference between 10s pointings and 100s pointings was marginal. Figure \ref{figit}b shows that observations with shorter pointings perform slightly better, however the integrated emission maps appear nearly identical.
We use the CASA recommended 10s pointing time as our default value,  but we suggest that the choice of pointing time may have little impact on actual observations.

\bibliography{converted_to_latex.bib}

\end{document}